\documentclass{aa}

\usepackage{graphicx}
\usepackage{txfonts}
\usepackage{subfigure}
\usepackage[dvips]{color}
\usepackage{natbib}
\bibpunct{(}{)}{;}{a}{}{,} 

\begin{document}

\title{Three-dimensional radiative transfer models of clumpy tori in Seyfert galaxies}

\author{M.~Schartmann
          \inst{1}\fnmsep\thanks{e-mail: schartmann@mpe.mpg.de}
          \and
          K.~Meisenheimer
          \inst{1}
          \and
          M.~Camenzind
          \inst{2}
          \and
          S.~Wolf
          \inst{1}
          \and
          K.~R.~W.~Tristram
          \inst{1}
          \and
          Th.~Henning
          \inst{1}}

\offprints{M.~Schartmann}

\institute{Max-Planck-Institut f\"ur Astronomie (MPIA), K\"onigstuhl 17, D-69117 Heidelberg, Germany
             \and 
           ZAH, Landessternwarte Heidelberg, K\"onigstuhl 12, D-69117 Heidelberg, Germany
           }

\date{Received  / Accepted }

\abstract{Tori of Active Galactic Nuclei (AGN) are made up of a mixture of hot and cold gas, as well as
 dust. In order to protect the dust grains from destruction by the surrounding hot gas as well as
 by the energetic (UV/optical) radiation from the accretion disk, the dust is
 often assumed to be distributed in clouds.}
{A new three-dimensional model of AGN dust tori is extensively investigated. 
 The torus is modelled as a wedge-shaped disk 
 within which dusty clouds are randomly distributed throughout the volume, by taking the dust density distribution 
 of the corresponding continuous model into account. We especially concentrate on the differences between clumpy and
 continuous models in terms of the temperature distributions, the surface brightness distributions and 
 interferometric visibilities, as 
 well as spectral energy distributions.}
{Radiative transfer calculations with the help of the three-dimensional Monte Carlo radiative transfer code MC3D 
 are used in order to simulate spectral energy distributions as well
 as surface brightness distributions at various wavelengths. In a second step, interferometric 
 visibilities for various inclination as 
 well as position angles and baselines are calculated, which can be used to
 directly compare our models to interferometric observations with the MIDI instrument.}
{We find that the radial temperature distributions of clumpy models possess significantly enhanced scatter compared
 to the continuous cases. Even at large distances,
 clouds can be heated directly by the central accretion disk. The existence of the silicate 10$\,\muup$m-feature 
 in absorption or in
 emission depends sensitively on the distribution, the size and optical depth of 
 clouds in the innermost part
 of the dust distribution. With this explanation, failure and success of previous 
 modelling efforts of clumpy tori can be understood. The main reason for this
 outcome are shadowing effects of clouds within 
 the central region. We underline this result with the help of several parameter 
 variations.
 After adapting the parameters of our clumpy standard model to the circumstances of the 
 Seyfert\,2 Circinus galaxy, 
 it can qualitatively explain recent 
 mid-infrared interferometric observations performed with MIDI, 
 as well as high resolution spectral data.}
{}

\maketitle

\section{Introduction and motivation}
\label{sec:intro}

According to today's knowledge, Active Galactic Nuclei (AGN) are powered
by accretion onto a supermassive black hole ($10^6-10^{10}\,M_{\odot}$,
e.g.\/ \citealp{Shankar_04}) residing in their centres. Thereby,  
gravitational energy is converted into heat by viscous processes within 
the surrounding accretion disk, which extends from the
marginally stable orbit up to several thousands of Schwarzschild radii. 
The emitted UV/optical light illuminates the attached, toroidally shaped dust
reservoir. 
 The concept of this {\it obscuring torus} was introduced in order to unify mainly two
 classes of observed spectral energy distributions (SEDs): 
 one shows a peak in the UV-region with overlayed broad and narrow optical emission lines, 
 the other class shows only narrow optical emission lines. 
 This can be interpreted as an inclination angle dependence. For
 viewing angles within the dust-free cone of the torus (type\,1 sources), direct signatures of the
 accretion disk (a peak in the UV-range) and the region close to the centre
 within the funnel of the torus show up. This is where gas moves fast and, therefore, produces
 broad emission lines (the region is hence called the {\it Broad Line
 Region (BLR)} of the nucleus). For edge-on lines of sight (type\,2 sources), the direct view onto the 
 centre is blocked and optical emission
 lines can only be detected from gas beyond the torus funnel. Being further away from
 the centre, it moves slower and hence produces narrow emission lines only.
 This is the so-called {\it Unified Scheme
 of Active Galactic Nuclei} \citep{Antonucci_93,Urry_95}. 
 First evidence for this scenario came from spectropolarimetric observations of
 type~2 sources \citep{Miller_83}, clearly displaying type~1
 signatures in the polarised light, which is scattered by electrons and
 tenuous dust within the funnel above the torus. 
 The opening angle of the torus can be estimated with the help of statistics of the different
 types of Seyfert galaxies. 
 \citet{Maiolino_95} find a ratio between Sy~2 to Sy~1 galaxies of
 4:1 in their sample, which results in an opening angle of the light cones of
 $74\degr$, in concordance with many observations of ionisation cones of 
 individual galaxies.
 Direct support for the idea of geometrically thick tori comes from recent
 interferometric observations in the mid-infrared \citep[e.g.\/][]{Jaffe_04,Tristram_07}.

 These tori are made up of at
 least three components: (i) hot ionised gas, (ii) warm molecular gas and (iii) dust.
 Krolik \& Begelman (1988) proposed that the dusty part has to be organised in a 
 clumpy structure in order to prevent the grains from being
 destroyed by the hot surrounding gas (with temperatures
 of the order of $10^6~$K) in which the clouds are supposed to be embedded.
 Another hint for the clumpy nature of the obscuring material -- in this case
 mainly for the distribution of neutral gas -- comes
 from X-ray measurements of the absorbing column density. 
\citet{Risaliti_02} claim that the observed 
 variability of these measurements 
 on timescales from months to several years can be explained by a clumpy
 structure of the torus.
 Combining X-ray absorbing column densities with spectral information further
 strengthens the claim for a clumpy distribution of the dust \citep{Shi_06}.  
 
Earlier work on torus simulations concentrated mostly on smooth dust
distributions 
\citep[e.g.\/][]{Pier_92b,Granato_94,Bemmel_03,Schartmann_05}.
This was mainly caused by the lack of appropriate (3D) 
radiative transfer codes and computational
power. 
Nevertheless, such models are good approximations for the case that 
the clumps that build up the torus are small compared to the total 
torus size, as is also shown in a parameter study described in 
Sect.~\ref{sec:volfill}.
These continuous models are able to
describe the gross observable features of these objects (see
\citealp[e.~g.][]{Schartmann_05}). However, problems arose from
too strong emission features of silicate dust compared to the observations, when
looking directly onto the
inner rims of the model structures (face-on views). They had never been
observed before that time, although almost all models showed them for the face-on
view. Therefore, much theoretical effort was undertaken in order to find models 
showing no silicate feature at all in the face-on case, while retaining the silicate absorption
feature in the edge-on case.  
\citet{Manske_98} for example succeeded in avoiding silicate emission features
with a flared dust disk of high optical depth in combination with an
anisotropic radiation characteristic of the central illuminating source.
A very promising
idea was to solve
the problem naturally by splitting the
dust distribution into single clouds.  
This was first attempted by \citet{Nenkova_02}. 
A one-dimensional code for the simulation of radiative transfer through
single clumps was used and, in a second step, the torus and its emitted
SED was assembled 
by adding many clouds of different aspect angles 
with the help of a statistical method. 
With this approach, they could show that a clumpy dust distribution of this kind can significantly
smear out the prominent silicate emission feature of the SEDs of type~1 objects 
at $10\,\muup$m for a large range of parameter values. No
more fine-tuning was needed, as in the previously proposed solutions with
the help of special continuous models. Subsequently, real 
two-dimensional radiative transfer calculations were undertaken by
\citet{Dullemond_05}. Clouds were modelled as concentric rings.  
A direct comparison between these kinds of clumpy models and the corresponding
continuous models did not show evidence for a systematic suppression of the
silicate feature in emission in the clumpy models.  

Meanwhile, silicate features
in emission were found with the help of the
Infrared Spectrograph (IRS) onboard the Spitzer space telescope 
\citep[e.~g.][]{Siebenmorgen_05,Hao_05,Sturm_05,Weedman_05}.
For these kinds of studies, Spitzer
is superior to other available facilities, due to its high
sensitivity and the coverage of a wavelength range including both silicate
features (at $9.7\,\muup$m and $18.5\,\muup$m) 
and the surrounding continuum emission. Silicate emission
features were found in different levels of AGN activity, 
ranging from very luminous quasars down to weak LINERS.
These findings are in good agreement with a geometrical unification
by an optically thick dusty torus, as silicate emission features can
be produced even in the simplest models.
But one has to be cautious, as due to the large beam of the Spitzer space telescope
and the low temperatures measured,
it is unclear whether these silicate features result from dust emission 
in the innermost parts of the torus
or from optically thin regions surrounding them. 
  
Very detailed simulations of clumpy tori were undertaken recently by
\citet{Hoenig_06}. They apply a similar method as \citet{Nenkova_02}, but use a
2D radiative transfer code for the simulation of SEDs of individual spherical clumps at
various positions in the torus and with various illumination patterns: directly
illuminated and/or illuminated by reemitted light of surrounding clouds. 
In a second step, these clouds are distributed according to physical models by 
\citet{Vollmer_04} and \citet{Beckert_04}. A comparison of the resulting SEDs 
and images with spectroscopic and 
interferometric observations shows good agreement. 
This model is characterised by a large number of small clouds with a very large
optical depth, especially close to 
the centre. We compare our models with these models in Sect.\,\ref{sec:torus_other}.

Despite the detection of geometrically thick dust tori in nearby Seyfert
galaxies (e.g.\/ \citealp{Jaffe_04,Tristram_07}), many questions remain: How are
these tori formed? How are they stabilised against gravity? Do steady torus
solutions exist? 
Several attempts to answer these questions have been made. For example
\citet{Krolik_88} and \citet{Beckert_04} support the scale-height of their
tori with the help of discrete clumps, moving at supersonic velocities,
maintained by mainly elastic collisions with the help of strong magnetic fields.
Other groups replace the torus by a magnetically-driven wind solution \citep{Koenigl_94}.
The most recent suggestion comes from \citet{Krolik_07}, building up on an idea of
\citet{Pier_92a}, where the scale-height of
tori can be maintained with the help of infrared radiation pressure, as shown
with an idealised analytical model. A more detailed review of possible solutions
and their drawbacks is given in \citet{Krolik_07}. 
Another possible scenario, where the effects of stellar feedback from a nuclear cluster 
play a major role, is discussed in \citet{Schartmann_08}. 

In this paper, we address the implications of clumpiness on the temperature
structure, the infrared spectral
energy distributions, surface brightness distributions as well as interferometric visibilities 
by implementing fully three-dimensional 
radiative transfer calculations through a clumpy dust distribution and discuss
the possible mechanisms causing this behaviour.
In Sect.\,\ref{sec:Model}, a description of our model is
given, before we present the basic results for our standard model (Sect.\,\ref{sec:results_stanmodel}) 
and for several parameter studies (Sect.\,\ref{sec:param_study})
and discuss the findings (Sect.\,\ref{sec:discussion}), as well as differences and
similarities to other models. In Sect.\,\ref{sec:MIDI_interferometry} we interpret our results in
terms of MIDI interferometric observations and compare them to data for the
Circinus galaxy. Finally we draw our conclusions in Sect.
\,\ref{sec:conclusions}. \\


\section{The model}
\label{sec:Model}

\subsection{Assembly of our clumpy standard model}
\label{sec:torus_assembly}

We apply a very simple, wedge-like torus geometry with a half opening angle of
$45\degr$ in order to gain
resolution. In our previous two-dimensional continuous {\it TTM}-models \citep{Schartmann_05}, 
the simulation of the whole
$\theta$-range was necessary, due to the radial as well as $\theta$-dependence of the
dust distribution. It resulted from an equilibrium between turbulent pressure forces and forces due to an
effective potential. The latter is mainly made up of gravitational forces due to the central black hole and the 
central stellar distribution, as well as rotation.
The cloudy dust distribution is set up on a spherical three-dimensional
grid $\vec{r}=(r,\theta,\phi)$, which is linear in $\theta$ and $\phi$ and logarithmic in $r$. 
To obtain the clumpy density structure,  
the following procedure is applied:
A random number generator (RAN2 taken from \citealp{Press_92})
determines the radial coordinate of the clump centre, which is equally distributed between the
inner and outer radius. The $\theta$ and $\phi$ coordinates are chosen such that
the resulting points are equally distributed on spherical shells. 
In a second step, the spatial distribution found so far is coupled to 
the dust density distribution of the continuous model:

\begin{eqnarray}
\begin{centering}
\label{equ:den_dis}
\rho_{\mathrm{cont}}(r,\theta,\phi) = \rho_0 \, \left(\frac{r}{1\,\mathrm{pc}}\right)^{\alpha}.
\end{centering}
\end{eqnarray}
  
The radii of individual clumps $a_{\mathrm{clump}}$ vary with distance from the centre according to a
distribution 
\begin{eqnarray}
\begin{centering}
a_{\mathrm{clump}} = a_0 \,\left(\frac{r_{\mathrm{clump}}}{1\,\mathrm{pc}}\right)^{\beta}.
\end{centering}
\end{eqnarray}
All cells within this clump radius are homogeneously filled with dust.
All clumps possess the same optical depth $\tau_{9.7\,\muup\mathrm{m}}^{\mathrm{clump}}$, 
measured along a radial ray through the clump centre. 
We further require that clumps always  
have to be completely contained within the model space, but are allowed 
to intersect. Such a combination of intersecting clumps will be called 
a {\it cloud} from now on. 
Thus, a cloud may contain overdensities, where the intersection happens.
A clump size distribution as described above seems to be reasonable, as
shear forces due to the differential rotation increase towards the centre.
Therefore, clouds are more easily disrupted in the inner part of the torus.
Furthermore, clouds become compressed when moving towards the centre due to the
increasing ambient pressure in a deeper potential well.

All other routines and algorithms used in this paper are identical to the
modelling described in \citet{Schartmann_05} and will only be mentioned briefly in
Sect.\,\ref{sec:preconditions}. 

\begin{figure}
  \centering
  \resizebox{0.8\hsize}{!}{\includegraphics[angle=0,clip]{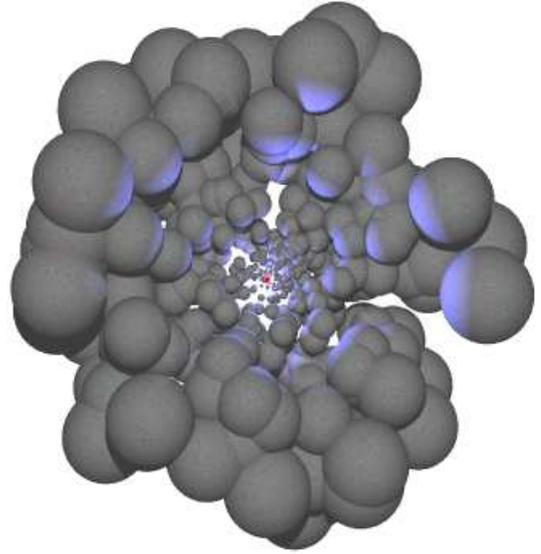}} 
  \caption[3D rendering of the clump distribution of our standard clumpy model]{3D
   rendering of the clump distribution of our standard clumpy torus model. 
  The chosen inclination angle corresponds to a Seyfert\,1 type (face-on) view onto the
  torus.}   
  \label{fig:standardmodel_vis} 
\end{figure}

The main model parameters of the continuous and clumpy distributions 
are summarised in Table\,\ref{tab:model_param}, 
\begin{table}
\centering
\caption[Model parameters for continuous and clumpy wedge models]{Main model 
  parameters for our continuous and clumpy standard model.}
\begin{tabular}{lcc}
{\bf both models} & & \\
\hline
\hline
inner radius of the torus & {\bf $R_{\mathrm{in}}$} & 0.4 pc \\ 
outer radius of the torus & $R_{\mathrm{out}}$ & 50 pc \\ 
half opening angle of the torus & {\bf $\theta_{\mathrm{open}}$} & $45\degr$  \\
total optical depth in equatorial plane 
 & {\bf $\left<\tau_{9.7\,\muup\mathrm{m}}^{\mathrm{equ}}\right>_{\phi}$} & 2.0 \\
exponent of continuous density distribution & {\bf $\alpha$}  & -0.5 \\
number of grid cells in $r$ direction & & 97 \\
number of grid cells in $\theta$ direction & & 31 \\
number of grid cells in $\phi$ direction & & 120 \\
 & & \\
{\bf additional in clumpy model} & & \\ 
\hline
\hline
number of clumps & {\bf $N_{\mathrm{clump}}$ } & 400 \\
exponent of clump size distribution & {\bf $\beta$}   &  1.0 \\
constant of clump size distribution & {\bf $a_0$} &  0.2 pc\\
optical depth of each clump & {\bf $\tau_{9.7\,\muup\mathrm{m}}^{\mathrm{clump}}$ } & 0.38 \\
average number of cells per clump & & 272 \\

\end{tabular}
\label{tab:model_param}
\end{table}
where the numerical values refer to our clumpy and continuous standard model.  
The dust density distribution for the clumpy case is shown in Fig.~\ref{fig:standardmodel_vis}. 
The torus possesses a volume filling factor of 30\% and the dust
mass was chosen such that the optical depth of the torus within the
equatorial plane (averaged over all angles $\phi$) reaches a value of two at
$9.7\,\muup$m. With this value, the resulting absorption column densities are
in concordance with observations of Seyfert type\,2 galaxies 
obtained with the IRS
spectrometer onboard Spitzer (see e.~g.~\citealp{Shi_06}), and the modelled silicate absorption 
feature depth compares well with observations. If not stated otherwise, the optical depth 
always refers to a wavelength of $9.7\,\muup$m throughout this paper.

As in \citet{Nenkova_02}, all clumps possess the same optical
depth in our standard model. A total optical depth within the equatorial plane
of $\left<\tau_{9.7\,\muup\mathrm{m}}^{\mathrm{equ}}\right>_{\phi}$ = 2.0
results in an optical depth of $\tau_{9.7\,\muup\mathrm{m}}^{\mathrm{clump}}=0.38$ along 
a radial ray through the centre of the clump. 
The corresponding continuous model has the same geometrical structure, continuously filled with dust according to the
density distribution given in equation\,\ref{equ:den_dis}.

\subsection{Summary of methods}
\label{sec:preconditions}

A very brief overview of the dust composition, the heating source and the numerical method of 
radiative transfer will be given in this section. 


Although several hints \citep{Maiolino_01b,Maiolino_01a,Jaffe_04} 
point to the possibility that dust in 
the nuclear regions of AGN is dominated by large grains, 
we will limit our 
present investigation to the classic MRN-model \citep{Mathis_77} 
for three reasons: first, and
most important, we aim for comparability with our earlier paper on 
continuous tori \citep{Schartmann_05}. Second, we have tested that our
essential results about the change in grain size distribution (Sect.\,3.9 in 
\citealp{Schartmann_05}) remain unchanged when distributing the dust in a clumpy structure.
Third, our approach, which explicitly takes into account
the size-dependent sublimation radius, is generically more robust against changes in the
grain distribution than calculations that ignore this effect.
For our current simulations, we represent the MRN-model by three different
grain species with 5 different grain sizes each. Taking different sublimation
radii of the various grains into account then partially accounts for the
destruction of small grains in the harsh environment of the quasar, 
as they possess larger sublimation radii. 
 

The dust distribution is heated by a point-like, central accretion disk 
with the SED of a mean
quasar spectrum (see Fig.\,3b in \citealp{Schartmann_05}). 
The radiation characteristic is chosen to follow a $|\cos(\theta)|$ law for
all wavelengths. For the simulations shown in this paper, the accretion disk SED
is normalised to a bolometric luminosity of $1.2 \times 10^{11}
\, L_{\sun}$, except for the comparison with the Circinus galaxy.


 \begin{figure}[t!]
   \centering
     \includegraphics[angle=0,clip,width=1.0\linewidth]{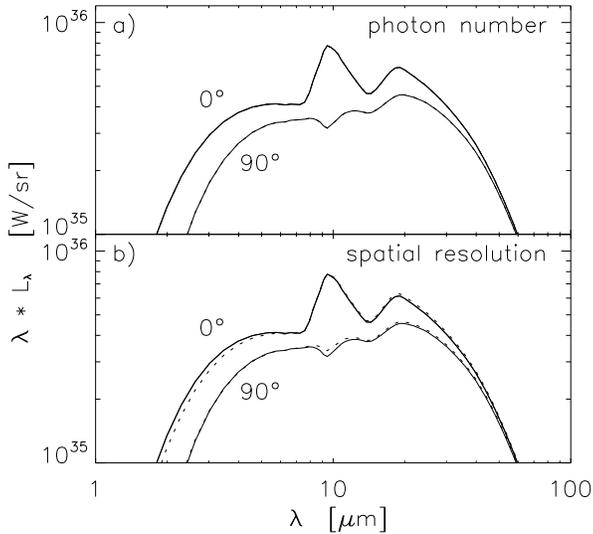} 
  \caption[Quality tests]{
     {\it a)} SEDs for a photon number study. The solid
     curves refer to our standard model and the dotted graphs (identical with the solid curves) 
     result after
     doubling the number of photon packages. {\it b)} SEDs for a resolution study:
     high resolution (solid curves -- our standard model) and a factor of 3
     reduced number of grid cells (dotted curves). Shown are the cases for inclination
     angles $0\degr$ and $90\degr$.}   
   \label{fig:quality_check}  
 \end{figure}

In order to obtain the temperature, the SEDs and the
surface brightness distributions of the dusty torus, we use the
three-dimensional radiative transfer code MC3D\footnote{MC3D ({\it Monte Carlo 3D}) has been tested
extensively against other radiative transfer codes for 2D structures \citep{Pascucci_04} and we also
performed a direct comparison for the special case of AGN dust tori with the
simulations of \citet{Granato_94}, one of the standard torus models for comparison,
calculated with his grid based code (see Fig.~4 in
\citealp{Schartmann_05}).} \citep{Wolf_99a,Wolf_03b}. We apply the
Monte Carlo procedure mainly for the calculation of temperature distributions and 
the scattering part whenever necessary,
whereas SEDs and surface brightness maps for dust
reemission are
obtained with the included raytracer. 
The main advantage compared to other codes is MC3D's capability to cope with real
three-dimensional dust density distributions, needed for a realistic modelling of
the dust
reemission from a clumpy torus. 
For this paper, we implemented the automatic determination of the sublimation surfaces of
the various grain species in
three dimensions. As we expect the sublimation to happen along irregularly
shaped surfaces in a three dimensional, discontinuous model, a raytracing
technique is used to solve the (1D) radiative transfer equation  
approximatively in all directions of the model space.   

For further information on the radiative transfer procedure used and the other
preconditions (mainly primary source and dust composition), see
\citet{Wolf_99a}, \citet{Wolf_00}, \citet{Wolf_01,Wolf_03b} and \citet{Schartmann_05}.

\subsection{Resolution study}
\label{sec:mc3d_restest}

In Fig.~\ref{fig:quality_check}a, we show SEDs for our standard
clumpy model (solid line, $5\cdot 10^6$ monochromatic photon packages) 
and for the same model, but with twice as many
photon packages ($10^7$) used for the simulation of the temperature distribution (dotted graphs, 
identical with solid lines). 
Despite slight differences in the temperature distributions of single grains, we find an almost
identical behaviour in the displayed SEDs, with differences smaller than the thickness of the lines.
Maps at 12\,$\mu$m display the same distribution with 
slight problems along the projected torus axis, which are not visible 
in the single surface brightness distributions
and without any noticeable effect on the interferometric visibility distributions calculated from these maps.  
In Fig.~\ref{fig:quality_check}b, the solid 
curves displays the SEDs for our high spatial resolution standard model and
the dotted lines refer to a model with a factor of roughly three less grid
cells. Only very small deviations are visible at short wavelengths.   
Fig.~\ref{fig:quality_check} clearly shows that the results and conclusions we draw from our simulations
are neither affected by photon noise nor by too low spatial resolution. 

  \begin{figure}[b!]
   \resizebox{1.0\hsize}{!}{\includegraphics[angle=0]{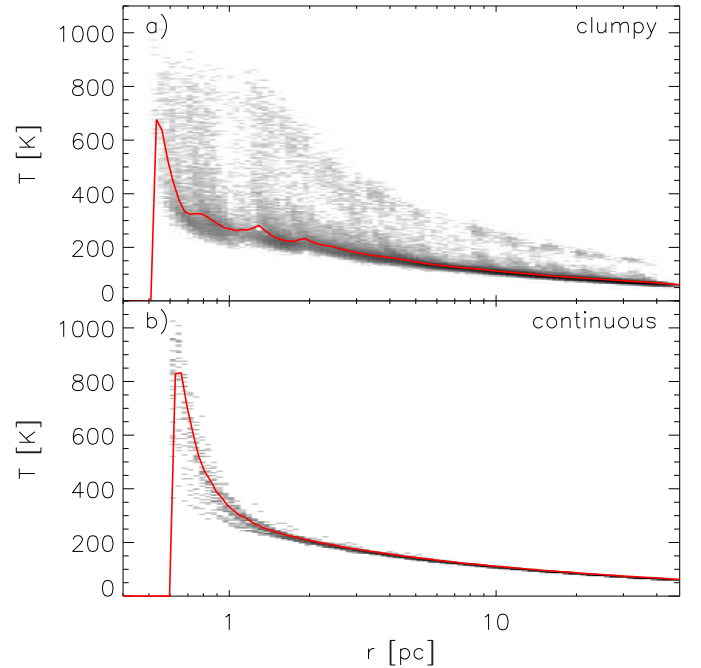}}
  \caption[Comparison between the radial temperature distribution of the
   clumpy model with the continuous model]{Comparison between radial temperature distributions 
     (for the smallest silicate grains) in all directions of the clumpy standard
     model (panel a) with the 
     temperature distribution of the corresponding
     continuous model (panel b). The red curve indicates the temperature averaged over
     all angles $\theta$ and $\phi$.}   
   \label{fig:temp_comp_equ} 
  \end{figure}

 
\section{Analysis of our standard model}
\label{sec:results_stanmodel}

In most of the SEDs discussed in this paper, only pure dust reemission SEDs are shown
and an azimuthal viewing angle of $45\degr$ is used, if not stated otherwise. 

\subsection{Temperature distribution}

In Fig.~\ref{fig:temp_comp_equ}, the temperature distribution of all cells in all $\theta$ and 
$\phi$ directions for the smallest silicate grain component is plotted a) for our 
clumpy standard model and b) for the corresponding
continuous model. The red curves show the radial temperature profile, averaged
over all $(\theta,\phi)$ directions. It is evident that the
spread of temperature values for a given distance from the primary 
radiation source is much larger for the clumpy models than for the
continuous ones.
Higher temperatures are possible even in parts of the torus further out, as dust
free or optically thin lines of sight exist far out, depending on the distribution 
of single clumps. Therefore, a direct illumination
of clouds is possible even at large radii.   
Concerning
the continuous model, the 
scatter decreases significantly from $2\,$pc outwards. Further in,  
the $\theta$ dependent radiation characteristic of the primary
source causes greater scatter due to higher temperatures further away from the
midplane.

\subsection{Viewing angle dependence}
\label{sec:viewing_angle}

\begin{figure}[b!]
  \centering
  \resizebox{\hsize}{!}{\includegraphics[angle=0]{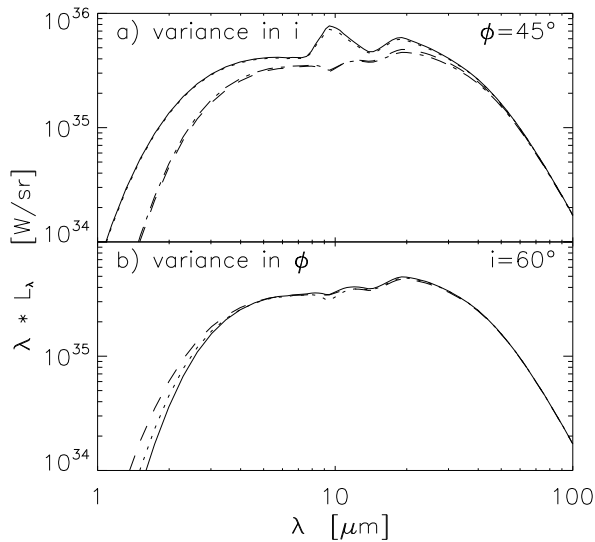}} 
  \caption[Viewing angle dependence of the SEDs of our clumpy standard
    model]{Dependence of 
    the SEDs on the viewing
    angle:  {\bf a)} different inclination angles for a common azimuthal angle
    $\phi$\,$=$\,$45\degr$. Inclination angles shown are:
    $0\degr$ (solid line), $30\degr$ (dotted line), $60\degr$ (dashed
    line), $90\degr$ (dashed-dotted line).  
    {\bf b)} Different azimuthal 
    angles $\phi$ for a common inclination angle of $i=60\degr$. Azimuthal angles shown are:
    $0\degr$ (solid line), $45\degr$ (dotted line) and $180\degr$ (dashed
    line).}  
  \label{fig:standardmodel_theta_dep_sed} 
\end{figure}

Fig.~\ref{fig:standardmodel_theta_dep_sed}a shows the dependence of the spectral
energy distributions on the inclination of the torus.
One can only see a clear distinction between lines of sight within the
dust-free funnel ($0\degr$ and
$30\degr$ inclinations) and those within the wedge-shaped disk ($60\degr$ 
and $90\degr$). This was already reported by \citet{Granato_94}, based on their
continuous wedge models. 
In our case, it is caused by the relatively large volume filling fraction and the large
clouds in the outer part of the torus. Therefore, only a weak dependence of the
dust density distribution on the polar angle exists, which we chose for simplicity. 
In our previous (2D) modelling (see \citealp{Schartmann_05}), 
we obtained the expected smooth transition in 
the polar direction.  
In Fig.~\ref{fig:standardmodel_theta_dep_sed}b, the azimuthal angle is
varied for a constant inclination angle of $60\degr$. 
Nearly identical SEDs result, which is understandable when considering our large volume
filling factors. The largest deviations appear at the shortest wavelengths, where 
the emission results from the hottest parts of the torus, which are also the most centrally 
concentrated parts. Therefore, this wavelength range is most sensitive to 
changes of the optical depth along the direct line of sight towards the centre. 

\begin{figure}
  \centering
  \resizebox{\hsize}{!}{\includegraphics[angle=0]{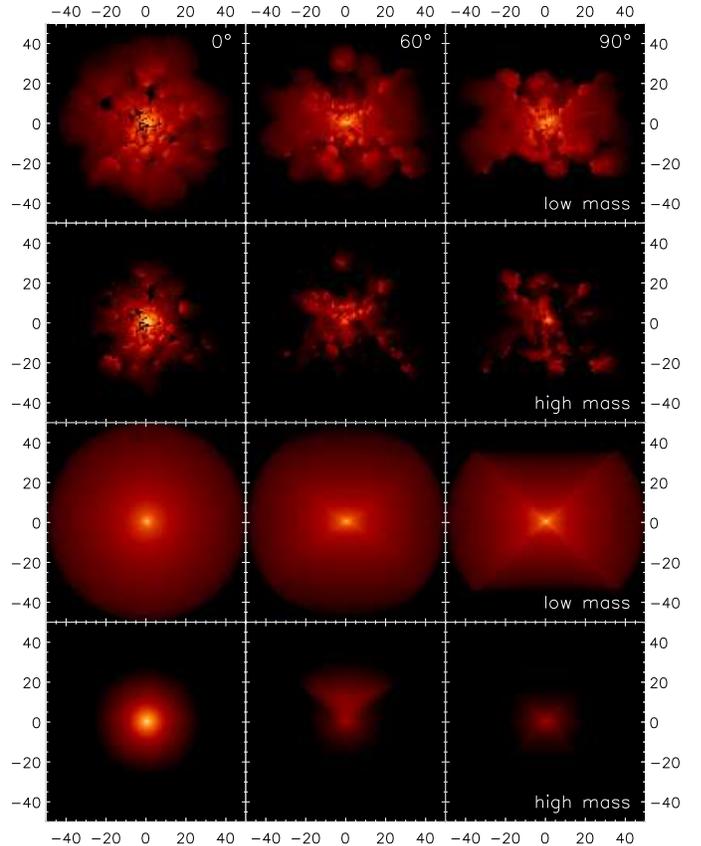}} 
  \caption[Comparison of images with various dust masses]{Inclination angle study of images of clumpy 
    (first two rows) and 
    continuous models (third and fourth row) with two different dust masses  
    at $12\,\muup$m. Shown are the extreme cases with half of the mass of the
    standard model (first and third row) and eight times the mass (second and fourth row). For details of the
    mass study, see Sect.~\ref{sec:dustmass_study}. 
    The inclination angles
    $i=0\degr$, $60\degr$, $90\degr$ are shown in different columns. The scaling is logarithmic with
    a range between the maximum value of all images and the $10^{-6}$th
    fraction of it (excluding the central point source). Labels are sizes in pc.}   
  \label{fig:dustmass_im} 
\end{figure}

The dependence on the inclination angle of images is shown in Fig.\,\ref{fig:dustmass_im} 
(upper two rows). 
It is especially interesting that the different inclination angles look  
very similar, which was not the case for the continuous model 
(see lower two rows of Fig.\,\ref{fig:dustmass_im}).
There, the images at larger inclination angles are dominated by the boundaries of 
the disk, which are not so well defined in the clumpy case. 
In the zoomed-in images 
(Fig.~\ref{fig:random_distribution}, upper row), the basic features of our
model are directly visible, as one can see the different illumination
patterns of clouds: Clouds in the innermost part are fully illuminated 
and, therefore, show bright inner rims and cold outer parts. Other clouds are
partly hidden behind clouds further in and appear as bright spots only.

\subsection{Wavelength dependency}
\label{sec:wavelength_dependency}

\begin{figure}
  \centering
  \resizebox{\hsize}{!}{\includegraphics[angle=0]{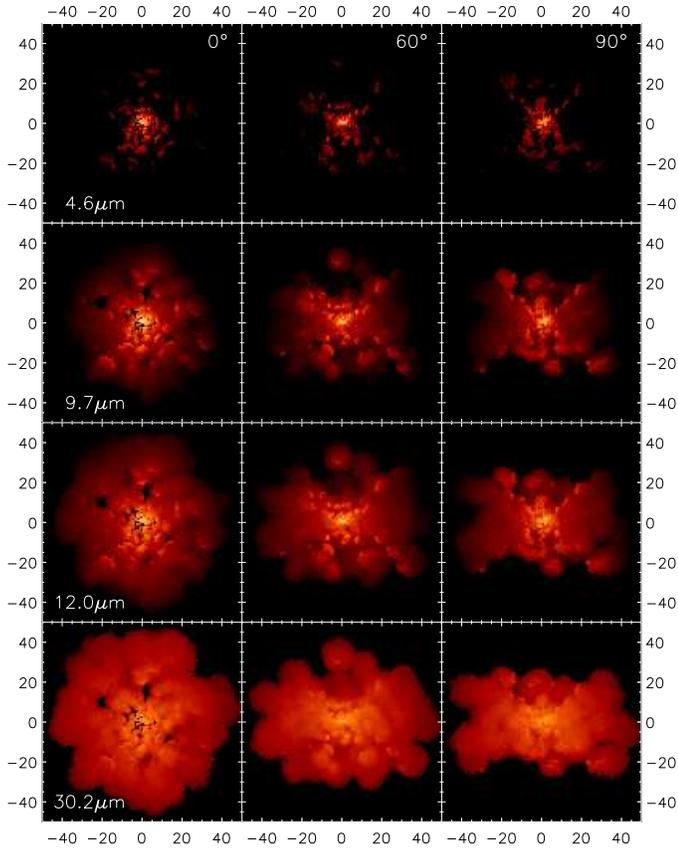}} 
  \caption[Wavelength dependence of the surface brightness distributions of our
  clumpy standard model]{Wavelength dependence of the surface brightness distributions:
    $\lambda=4.6\,\muup$m (upper row), 
    $\lambda=9.7\,\muup$m (second row), $\lambda=12.0\,\muup$m (third row) and
    $\lambda=30.2\,\muup$m (lower row). 
    Within the rows, the inclination angle changes from face-on view (leftmost
    panel) over $60\degr$ to $90\degr$ (rightmost panel).
    The images are given in
    logarithmic scaling with a range of values between the
    global maximum of all images and the
    $10^{-5}\mathrm{th}$ fraction of it (central source excluded). Labels are in pc.}
  \label{fig:standardmodel_sb_wave_dep} 
\end{figure}

Fig.~\ref{fig:standardmodel_sb_wave_dep} shows the wavelength dependency of
our standard model. At short wavelengths, the hottest inner parts
dominate the brightness distribution. 
Further out, a few more directly illuminated clumps are
visible as bright spots.
At the longest wavelengths, emission arises from clumps all over the
torus, as colder dust emits strongest at these wavelengths. This dust is spread
over a larger volume, due to the steeply decreasing temperature distribution
at small radii. 
Furthermore, the extinction
curve has dropped by a large factor at these wavelengths and, therefore, the torus becomes
optically thin and the whole range of cloud sizes is visible.  

\section{Parameter variations}
\label{sec:param_study}

\subsection{Different realisations of the clumpy distribution}

As already discussed in \citet{Schartmann_05}, SEDs of dust reemission depend strongly on the distribution of
dust in the innermost region. Changing the random arrangements of clumps -- as
done in this section -- therefore is expected to cause significant changes of the
SEDs, especially for the case of a small number of clouds. The second
important parameter is the optical depth of the single clumps. 
The larger it is, the stronger is the
dependence of the SEDs on the dust distribution in the innermost region. 
For the case of our modelling, the small number of clouds is expected to 
cause large differences in the observed SEDs. But this effect is partially compensated --
in most of the simulations -- by optically thin individual clumps, resulting in a more similar behaviour
of the SEDs.
 
\begin{figure*}
  \centering
  \resizebox{0.6\hsize}{!}{\includegraphics[angle=0]{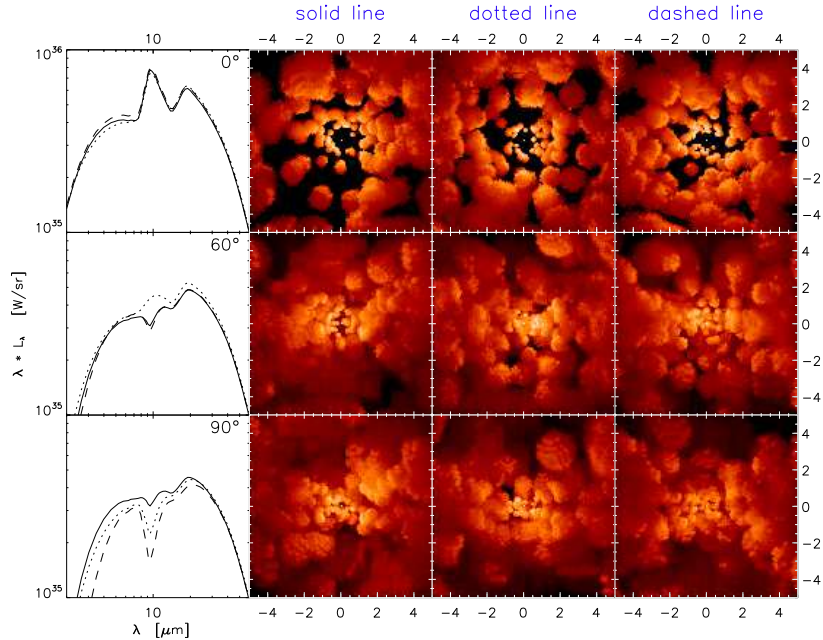}} 
  \caption[Dependence on the random arrangements of clumps]{Different random
           arrangements of 
           clumps. The rows show three different inclination angles: 
           $0\degr$ (first row), $60\degr$ (second row), $90\degr$ (third row). Given in columns are 
           the SEDs (first column) and the images at $12\,\muup$m for the three
           different random cloud arrangements (column 2 to 4), all having the
           same parameters (see Table~\ref{tab:model_param}). Here,
           the solid line in the SED corresponds to the first column of images, the
           dotted line to the second and the dashed line to the third column.
           Images 
           are given in logarithmic color scale ranging from the maximum of
           all images to the $10^{-4}$th part of it (excluding the central source).
           Labels denote distance to the centre in pc.}  
  \label{fig:random_distribution} 
\end{figure*}

Looking at the simulated SEDs 
and matching them with the images, 
the following results can be seen (compare to Fig.~\ref{fig:random_distribution}):

At $0\degr$ inclination angle (upper row), we observe nearly identical 
SEDs. The dashed line (corresponding to the fourth column) shows a slightly
enhanced flux at short wavelengths,
as a larger number of clouds are close to the
central source. In the third column, 
the cloud number density in the central part is the lowest of the three examples. 

For the case of the middle row ($i=60\degr$), the largest deviations are visible for the case 
of the dotted line (third column). Here, the silicate feature even appears in emission.
This is visible in the surface brightness distributions, as more directly illuminated clouds 
are visible on unobscured lines of sight, resulting in a brighter central region compared to the other two maps.
At an inclination angle of $90\degr$ (third row), absorption along the line of sight increases 
drastically from the second to the fourth column, visible in a deepening of the silicate absorption 
feature and the darkening of the central region of the surface brightness distributions.

\subsection{Different volume filling factors}
\label{sec:volfill}

Starting
from the standard model with a volume filling factor of $30\%$ and 400 clumps
within the whole model space, we halved it once by distributing only 160
clumps within the calculation domain and doubled it, for which 1500 clumps were
needed due to the applied procedure of randomly distributing clumps.

\begin{figure}[b!]
  \centering
  \resizebox{\hsize}{!}{\includegraphics[angle=0]{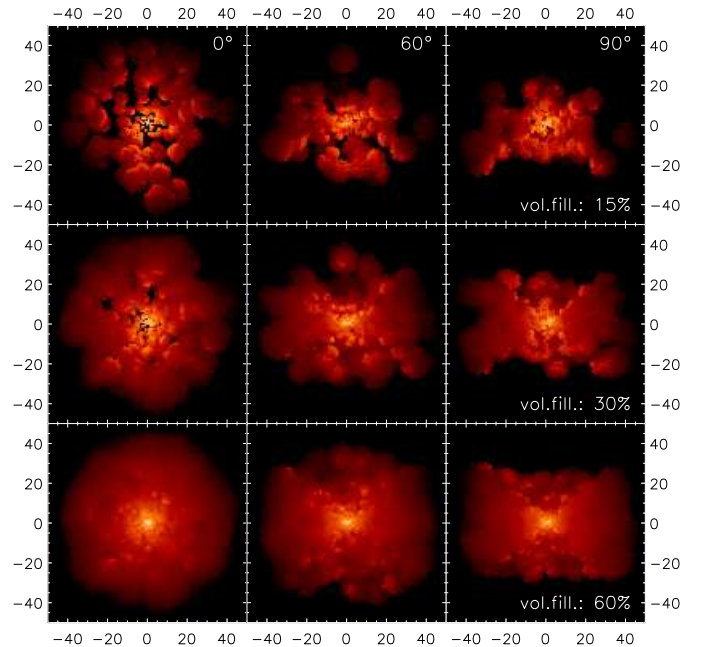}} 
  \caption[Surface brightness distributions at various volume filling factors
    of our clumpy standard model]{Different volume filling factors: 
    $15\%$ (upper row), $30\%$ (middle
    row), $60\%$ (lower row) for $\lambda=12.0\,\muup$m. 
    From left to right, the inclination angle
    changes: $i=0\degr$, $60\degr$, $90\degr$. The scaling is logarithmic with
    a range between the maximum value of all images and the $10^{-6}$th
    fraction of it (excluding the central point source). Length scales are given in pc.} 
  \label{fig:volfill_sb} 
\end{figure}
\begin{figure}[b!]
  \centering
  \resizebox{0.75\hsize}{!}{\includegraphics[angle=0]{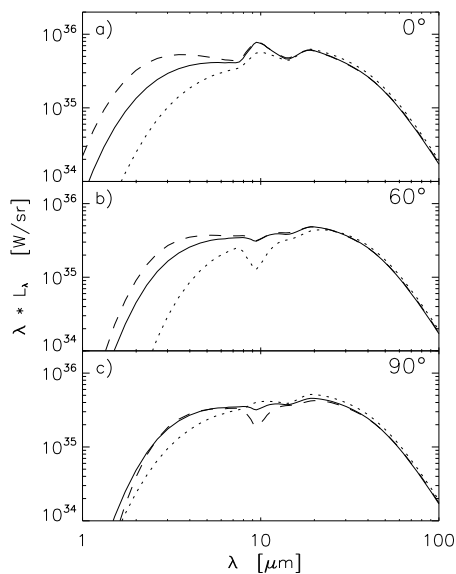}} 
  \caption[SEDs at various volume filling factors
    of our clumpy standard model]{Different volume filling factors: 
    $15\%$ (dotted line), $30\%$ (solid
    line), $60\%$ (dashed line). 
    Rows show SEDs for the three inclination angles: $i=0\degr$, $60\degr$, $90\degr$. The
    viewing angle in $\phi$-direction is always $45\degr$.}  
  \label{fig:volfill_sed} 
\end{figure}

The resulting surface brightness distributions at $\lambda\,=\,12.0\,\muup$m are 
shown in Fig.~\ref{fig:volfill_sb}
with the three models given in different rows and for three different
inclination angles: $i=0\degr$, $60\degr$, $90\degr$.
In the case of the lowest volume filling factor, individual clouds are
visible. The distribution of the surface brightness of these individual clouds
reflects the temperature structure within single clumps. The directly
illuminated clumps are hotter and, therefore, appear brighter. When adding
more and more clumps (increasing the volume filling factor), the chance of
directly illuminating clumps further out decreases and at higher filling factors
it is only possible for clumps close to the funnel. This is clearly visible at
higher inclination angles: the higher the volume filling factor, the clearer
the x-shaped feature appears, as only clumps within or close to the funnel can
be directly illuminated. At a volume filling factor of $60\%$, the surface
brightness distribution looks very similar to that of the corresponding
continuous model (compare to Fig.~\ref{fig:dustmass_im}). 
For large volume filling factors and close to edge-on, substructure is only visible from
clouds in a viewing direction towards the dust-free cones. 

The corresponding SEDs are shown in Fig.~\ref{fig:volfill_sed}. 
With increasing filling factor, more and more flux at short wavelengths
appears for the face-on case, as seen in \citet{Schartmann_05}.
The shape of the clumpy model SEDs resemble the
corresponding continuous model most (compare to Fig.~\ref{fig:dustmass}, right column) for
the highest volume filling factor. 
Concerning the silicate feature, it increases slightly in
emission as the amount of dust at the appropriate temperature increases as
well (visible at the transition from the lowest to the medium volume filling factor). 
The silicate absorption feature at higher inclinations strongly depends on the viewing
angle (compare Fig.~\ref{fig:volfill_sed}b and c) especially for the model with 
the least number of clumps (dotted line).    
Thus, this study shows the validity of the simplification of using a smooth 
dust distribution in the case of very high torus volume filling factors, as was assumed in previous
simulations.

\subsection{Dust mass study}
\label{sec:dustmass_study}

To study the dependence of the SEDs on the 
optical depth of the torus, we carried out a study with
0.5, 1, 2, 4 and 8 times the dust mass in the standard model. 
This leads to an optical
depth at 9.7$\,\muup$m within the equatorial plane, averaged over all angles of
$\phi$ of
$\left<\tau_{9.7\,\muup\mathrm{m}}^{\mathrm{equ}}\right>_{\phi} = $ 1, 2, 4, 8, 16.
Single clumps then change from optically thin to optically thick
($\tau_{9.7\,\muup\mathrm{m}}^{\mathrm{clump}} = 0.19, 0.38, 0.76, 1.52, 3.04$).
\begin{figure*}
 \centering
  \resizebox{0.6\hsize}{!}{\includegraphics[angle=0]{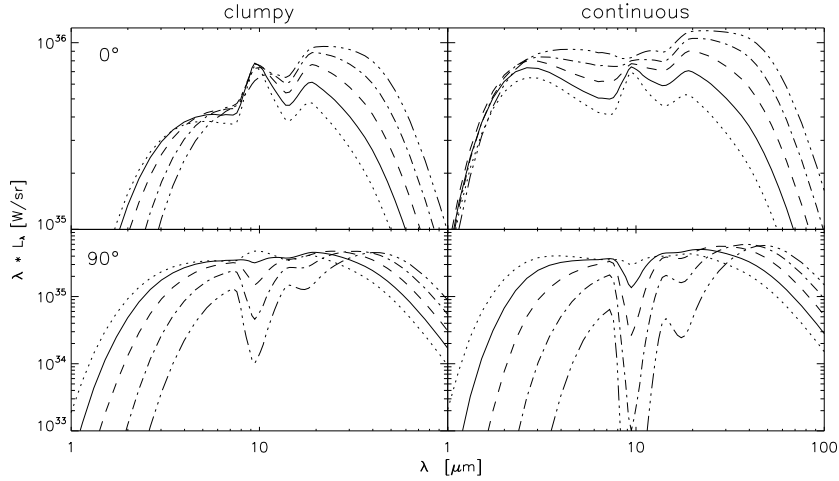}} 
 \caption[SEDs for different enclosed dust masses]{SEDs for different enclosed dust masses. The left
    column shows the
    case for the clumpy models and a face-on view (upper row) as well as an edge-on
    view (lower row).  In the right column, continuous models are displayed. 
    The solid line corresponds to the standard model, the dotted to half of the mass, 
    the dashed double the mass, the dash-dotted to four times the mass and the
    dash-triple-dotted to eight times the mass of the standard model. \label{fig:dustmass}} 
\end{figure*}

\begin{figure}
  \centering
  \resizebox{0.75\hsize}{!}{\includegraphics[angle=0]{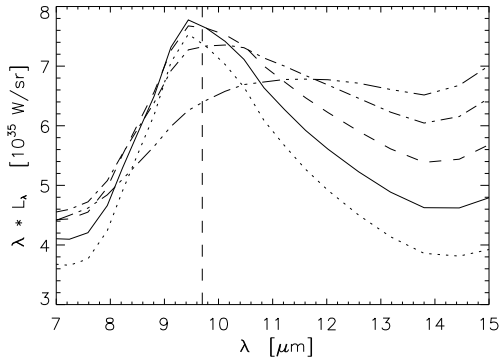}} 
  \caption[Close-up of the spectrum between 7 and $15\,\muup$m for the face-on
  case of the dust mass study]{Close-up 
  of the spectrum between 7 and $15\,\muup$m of the face-on case of the dust mass study for the clumpy
  model in linear display. The lines are defined as in Fig.~\ref{fig:dustmass}.}   
  \label{fig:dustmass_close} 
\end{figure}

The resulting behaviour of the SEDs is shown in
Fig.~\ref{fig:dustmass}, where it is also compared to the corresponding
continuous models. Concerning the silicate feature (in emission) for the
face-on case (top row), we see a very similar behaviour of the SEDs of the clumpy and
continuous model. Increasing the mass and with it the total optical depth leads to a flattening of
the SED around the silicate feature, even more pronounced in the continuous case. 
In addition to that, a slight shift of the maximum of the
silicate feature towards longer wavelengths is visible for the case of the highest dust
mass, apparent in the zoom-in of Fig.\,\ref{fig:dustmass} around the silicate feature 
(see Fig.~\ref{fig:dustmass_close}). 
This is due to the increasing underlying 
continuum towards longer wavelengths.
Although the principal behaviour of the silicate feature is identical for our
clumpy and our continuous model, the reasons
differ: In the case of a continuous wedge-like torus, the inner, directly illuminated 
walls are only visible through a small amount of dust. From step to
step, the walls become more opaque and shield the directly illuminated 
inner rim better,
decreasing the height of the silicate feature. 
This was not the case for
our continuous {\it TTM}-models in \citet{Schartmann_05}. With them, it was not possible
to significantly reduce the silicate feature height within reasonable optical depth ranges. 
This was caused by the fully visible, directly illuminated
inner funnel. 
Therefore, 
in the wedge-shaped continuous models,
the reduction of the feature is an artefact caused by the unphysical, purely geometrically 
motivated shape. Furthermore, it also involves very deep and so far unobserved 
silicate absorption features for 
the edge-on case (see lower right panel in Fig.~\ref{fig:dustmass}).
Concerning the clumpy model, the explanation for the flattening of the
silicate feature in the edge-on case with increasing dust mass of the torus will be given in
Sect.~\ref{sec:model_reduction}. 
    
The Wien branches show different behaviour when looking face-on.
For the case of the clumpy torus model, increasing
the optical depth means that the Wien branch moves to larger
wavelengths, as expected for the edge-on case. This is understandable when most of the directly
illuminated surfaces of the clouds are then hidden behind other clouds, an argument which
is not valid if the clouds are too optically thin in the inner part.\\
For the edge-on case, we qualitatively obtain a comparable behaviour as in the
continuous case, because of the large number of clumps and the same
optical depth within the equatorial plane. 
But a very important difference can be seen in the appearance of the silicate
feature in absorption:
when we want to have only very weak silicate emission features in the
face-on case, a large optical depth is needed, resulting in an unphysically
deep silicate feature in absorption in the edge-on case of the continuous models, 
whereas the silicate feature remains
moderate for many lines of sight for the clumpy model, where we see a large
scatter for different random arrangements of clumps (compare to Fig.~\ref{fig:random_distribution}). 

Concerning surface brightness distributions (see
Fig.~\ref{fig:dustmass_im}), one can see that
the objects appear smaller at mid-infrared wavelengths for the case of higher dust
masses: the larger the optical
depth, the brighter the inner region and the dimmer the outer part.
This is caused by a steepening of the radial dust temperature distribution with
increasing mass of the objects, as the probability of
photon absorption increases in the central region. Especially for the
continuous case, the asymmetry at intermediate
inclination angles becomes visible for larger optical depths caused by extinction on the line of sight due to 
cold dust in the outer parts of the torus.

\subsection{Concentration of clumps in radial direction}

\begin{figure}[t!]
  \centering
  \resizebox{\hsize}{!}{\includegraphics[angle=0]{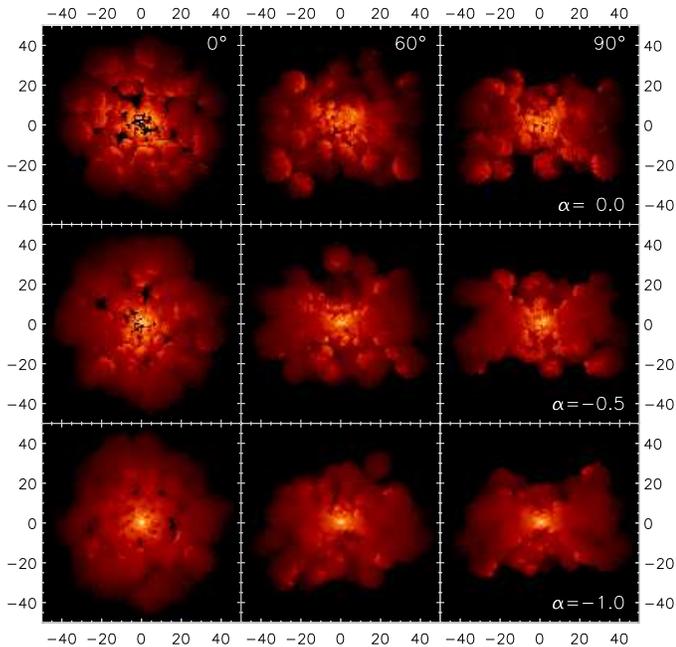}} 
  \caption[Surface brightness distributions for various slopes of the density
    distribution]{Surface brightness distributions for various slopes of 
    the density distribution in the corresponding continuous
    model ($\rho_{\mathrm{cont}} \propto r^{\alpha}$)
    leading to different concentrations of clumps in the radial
    direction in the clumpy model. The slopes are: $\alpha=0.0$ (upper row), $\alpha=-0.5$ (middle
    row), $\alpha=-1.0$ (lower row). From left to right, the inclination angle
    increases from face-on to edge-on: $i=0\degr$, $60\degr$,
    $90\degr$. Shown are images at $12\,\muup$m with a logarithmic color scale
    ranging from the maximum of all images to the $10^{-6}$th part of it.
    Labels denote the distance to the centre in pc.}  
  \label{fig:concentration_sb} 
\end{figure}

As already described in the model section (\ref{sec:Model}), clump positions
are also chosen in accordance with the density distribution of the
corresponding continuous model. Therefore, changing the slope of this radial density
distribution, defined to be 
$\rho_{\mathrm{cont}}(r,\theta,\phi) = \rho_0 \, \left(\frac{r}{1\,\mathrm{pc}}\right)^{\alpha}$, 
leads to a different concentration of clumps along radial
rays. In this section, we vary the slope of the distribution $\alpha$ from a
homogeneous dust distribution ($\alpha=0.0$) over $\alpha=-0.5$ (our standard model) to $\alpha = -1.0$.
Decreasing $\alpha$ leads to an enhancement of the clump number density towards the
central region. In order to keep the volume filling fraction at a constant
level of $30\%$, we need to increase the number of clumps, as
their size decreases towards the central region. All clumps possess the same
optical depth. In order to have a constant mean optical
depth in the midplane,
the total dust mass has to be decreased. For an overview of the modified parameters see Table\,
\ref{tab:slope}.
\begin{table}[!b]
\caption[Varied parameters of the clump concentration study]{Varied parameters of the clump concentration study.}
\centering
\begin{tabular}{lccc}
\hline
\hline
Parameter & $\alpha=0$ & $\alpha=-0.5$ & $\alpha=-1$ \\
\hline
No. of clumps & 250 & 400 & 900 \\
Dust mass [$\mathrm{M}_{\sun}$] & 22950 & 12562 & 6418 \\
\hline
\end{tabular}
\label{tab:slope}
\end{table}
The change of clump concentration can be seen directly from the simulated images at $12.0\,\muup$m
in Fig.~\ref{fig:concentration_sb}, especially in the face-on case (first column). In the
upper panel, single reemitting clumps are visible in the central region. This
changes more and more to a continuous emission for the case of the highest
cloud concentration in the centre due to multiple clumps along the line of
sight and intersecting clumps. At higher inclination angles, the higher
concentration leads to a sharper peak of the surface brightness. 

\begin{figure}[ht]
\centering
  \resizebox{0.95\hsize}{!}{\includegraphics[angle=0]{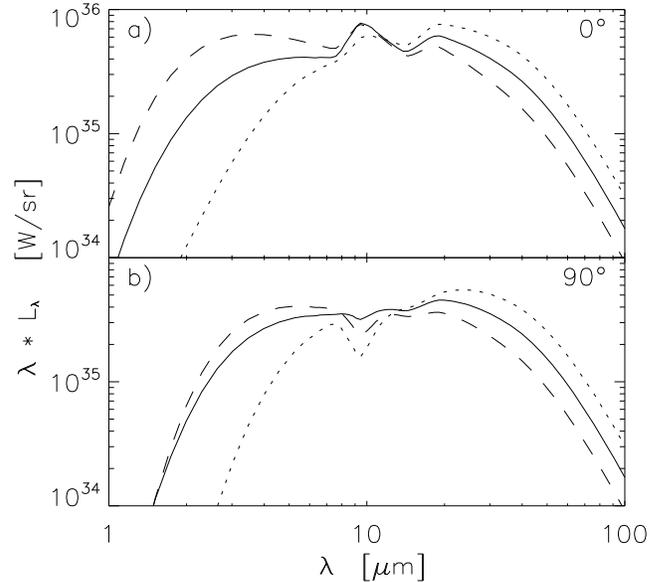}}
  \caption[SEDs for the clump concentration study]
    {SEDs for the clump concentration study. The varied slopes of the underlying density distribution are: 
    $\alpha=0.0$ (dotted line), $\alpha=-0.5$ (solid
    line), $\alpha=-1.0$ (dashed line):
    {\bf a)} face-on,
    {\bf b)} edge-on.} 
  \label{fig:concentration_sed}
\end{figure}

The same behaviour is visible in the corresponding SEDs
shown in Fig.~\ref{fig:concentration_sed}. Decreasing the amount
of dust in the centre near the heating source leads to decreasing flux at
near-infrared wavelengths, whereas the flux at far-infrared wavelengths
increases (reflecting the enhancement of dust in the outer part). 

\subsection{Dependence on the clump size distribution}

In our clumpy standard model, a radially changing clump size 
proportional to the radial distance to the
centre was chosen. 
In this section, we test the effects of decreasing the slope
$\beta$ 
of the radial size distribution $a_{\mathrm{clump}} = a_0
\,\left(\frac{r_{\mathrm{clump}}}{1\,\mathrm{pc}}\right)^{\beta}$
of the clumps. This is done in a way that the volume filling fraction as well
as the optical depth in the midplane, averaged over all azimuthal angles $\phi$, remain
constant. It is achieved by changing the proportionalisation constant of the
clump size distribution $a_0$ and the total dust mass of
the torus. 
Doing this results in very well resolved clumps in the inner part. Beyond a
distance of approximately $25\,$pc, the number of grid cells per clump drops below the value
of our standard model.

\begin{figure}[b!]
   \resizebox{1.0\hsize}{!}{\includegraphics[angle=0]{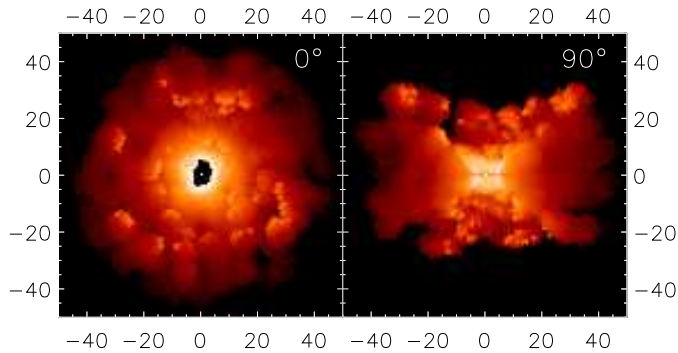}} 
  \caption[Images for a clumpy distribution with a constant clump size]{Images 
    at $12\,\muup$m and inclination angles $i=0\degr$ and
    $90\degr$ for a clumpy distribution with a constant clump size, independent
    of the radial position. The scaling is logarithmic with
    a range between the maximum value of all images and the $10^{-6}$th
    fraction of it (excluding the central point source).}  
  \label{fig:clump_size_sb} 
\end{figure}

Surface brightness distributions for the extreme case of a constant clump size are
shown in Fig.~\ref{fig:clump_size_sb}. Due to the large clump radius of $5\,$pc
even in the central region and that
fractions of clumps at the model space boarder are prohibited, it leads to a density
distribution with a quite large, unevenly shaped central cavity, 
as can be seen in the face-on view (left panel of Fig.~\ref{fig:clump_size_sb}). 
The inner rim is given by only a few
intersecting clumps, instead of the otherwise defined spherical central
cavity. Therefore, in the edge-on case, the surface brightness distribution
shows an inner boundary, which is bent towards the centre (convex shaped). 
In these models, due to the large clump
size in the inner region, many clumps intersect, producing a nearly continuous
dust distribution at the inner boundaries, which lets the -- typical for continuous
models -- x-shaped structure appear again. For the same reason, the
extinction band due to the $|\cos(\theta)|$-radiation characteristic is visible in
the edge-on view. Especially at the $90\degr$ inclination angle,
single clumps are directly visible (above and below the centre). 
In these cases, their shading
directly shows the illumination pattern due to the primary source (accretion
disk), emission from other clumps and extinction from the foreground dust
distributions.  

\begin{figure}
  \centering
 \resizebox{0.95\hsize}{!}{\includegraphics[angle=0]{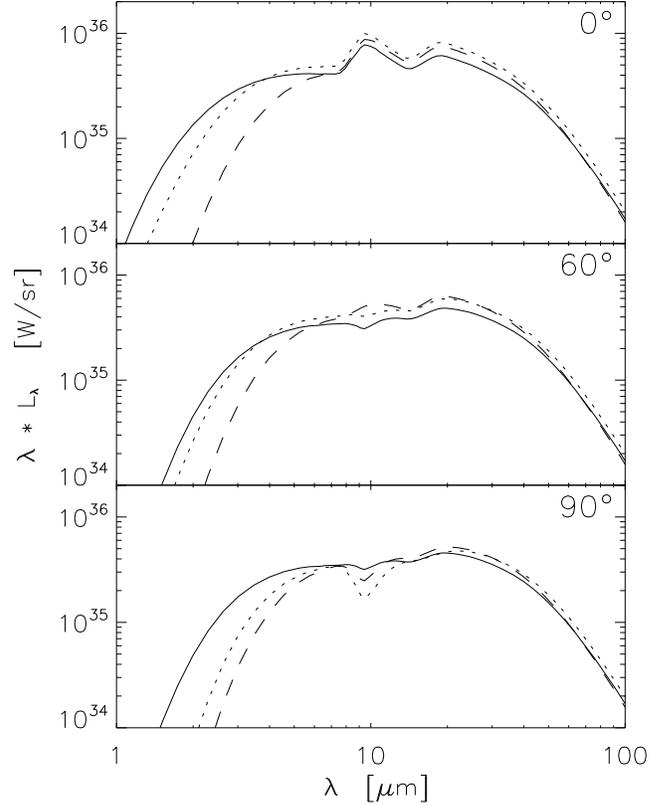}} 
  \caption[Dependence of SEDs on the clump size for different inclination
    angles]{Dependence 
    on the clump size for different inclination angles
    (rows: $0\degr$, $60\degr$, $90\degr$). The solid line corresponds to our standard model with
    $\beta = 1$. For the dashed line,
    clumps have equal size ($a_{\mathrm{clump}} = 5\,$pc, $\beta=0$), independent of the
    radial position and the dotted line corresponds to an intermediate model
    with $\beta=0.5$.}  
  \label{fig:clump_size_sed} 
\end{figure}

The corresponding SEDs (Fig.~\ref{fig:clump_size_sed}) mainly reflect the
increase of the inner cavity and, therefore, the lack of flux at short wavelengths. The
convex shape of that region causes a larger directly illuminated area at the
funnel walls and, therefore, slightly strengthens the silicate emission
feature in a face-on view. A different appearance (emission/absorption)
of the $10\,\muup$m feature at $i=60\degr$ (middle panel) is seen. This is due to the
lower number density of clumps in the inner part, enforced by the restriction of
having only whole clumps within the model space.

A dust mass study for the case of the large, 
constant diameter clump model ($\beta=0$) reveals the same behaviour as discussed in 
Sect.~\ref{sec:dustmass_study} when looking edge-on onto the torus. 
However, the face-on case differs:
only the relative height of the silicate feature changes slightly. This was already
observed in our {\it TTM}-models in \citet{Schartmann_05} and is due to the
now inwardly bent inner walls of the funnel (see Fig.\,\ref{fig:clump_size_sb}, right panel), 
caused here by the very large and
spherical clumps in the innermost torus region.

\section{Discussion}
\label{sec:discussion}

\subsection{Explanation for the reduction of the silicate feature}
\label{sec:model_reduction}

\begin{figure}[b!]
  \centering
  \resizebox{0.75\hsize}{!}{\includegraphics[]{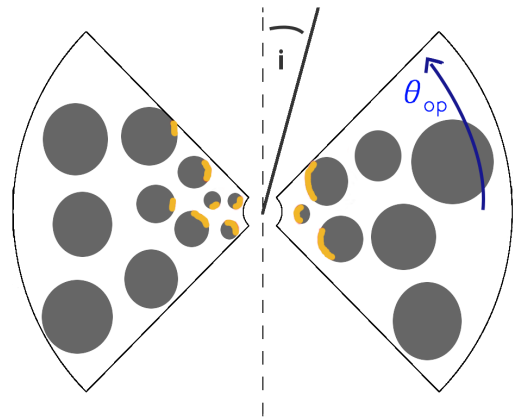}} 
  \caption[Sketch of our clumpy torus model]{Sketch of our clumpy torus model. 
  Indicated in yellow are directly 
  illuminated surfaces of the clumps. 
  $i$ is the inclination angle, $\theta_{\mathrm{open}}$ is the half 
  opening angle of the torus.}   
  \label{fig:explain_model} 
\end{figure}

The results shown in the subsections before can be explained with the following
model, which was partially discussed by \citet{Nenkova_02}. 
It is illustrated in Fig.~\ref{fig:explain_model},
where yellow denotes directly illuminated clump surfaces.  
Many of the explained features can also be seen
in the zoomed-in versions of the surface brightness distributions 
(Fig.~\ref{fig:random_distribution}) for the face-on case (upper row).

As already pointed out in \citet{Schartmann_05}, the SEDs of dust tori in the mid-infrared
wavelength range are
mainly determined by the inner few parsecs of the toroidal dust
distribution. In each of the central clouds of the clumpy model, 
the dust temperature drops from
the inner directly illuminated edge towards the cloud's outer surface.  

With an inclination angle close to
$i=90\degr$, we expect -- for realistic volume filling factors --
comparable behaviour of the SED as in the continuous case. The silicate feature has a smaller 
depth, as discussed in Sect.\,\ref{sec:dustmass_study}. But the situation changes with decreasing
inclination angle. Here, one has to distinguish between different cases: 

\begin{enumerate}
\item With a relatively high volume filling factor and a not too small
  extension of the clouds in the central region of the torus, it is likely
  that the directly illuminated part of most of the clouds is hidden from direct
  view by other clouds. Therefore, the directly illuminated surface area
  is reduced compared to the corresponding continuous
  model. As this area is responsible for the emission fraction of the silicate
  feature within the SED, it shows less silicate emission. 
  In order to produce such a shadowing effect, clouds have to
  possess a large enough optical depth, which means that they have to be 
  either small or massive ($\tau_{\mathrm{clump}} \propto
  m_{\mathrm{clump}}\,a_{\mathrm{clump}}^{-2}$, where $m_{\mathrm{clump}}$ is
  the mass of the clump). 

\item For the case of a too small optical depth of the clumps in the innermost
  region, we expect the silicate feature to appear in strong emission.

\item Another possibility of producing silicate features in emission is when the model
  possesses a small number of clumps in the inner part, making the
  shadowing effect inefficient. 

\item A further effect on the silicate feature strength 
  arises from the grain dependent sublimation implemented 
  in our models. As graphite grains possess a higher sublimation temperature as 
  silicate grains, they are able to partially shelter the silicate grains from 
  direct irradiation. 
\end{enumerate}

Thus, the strength of the silicate feature is mainly determined by
the distribution, size and optical depth of the clouds in the direct vicinity
of the sublimation surface of the dust. 
We will see in the next section that this finding well explains 
the fact that \citet{Nenkova_02}
and \citet{Dullemond_05} come to different conclusions concerning the reduction of the
strength of the silicate feature due to clumpiness, as the distribution (and
size) of their clouds differ.

\subsection{Comparison with other torus models}
\label{sec:torus_other}

The results of \citet{Nenkova_02} - the pioneering work in the field of clumpy tori - 
are broadly consistent with the explanations given in Sect.\,\ref{sec:model_reduction} 
of this paper. 

\citet{Dullemond_05} model 2D clumps in the form of rings with a two-dimensional 
radiative transfer code. In contradiction to all other simulations, no systematic 
reduction of the silicate feature due to clumpiness is found. The reason for this is understandable 
with the explanations given in Sect.\,\ref{sec:model_reduction}, as their model features 
a small clump number density in the central region and  
shadowing effects are rather small. Therefore, they find both strengthening of the 
silicate feature and reduction, depending on the random ring distribution.
 
\begin{figure}[b!]
  \centering
  \resizebox{0.95\hsize}{!}{\includegraphics[]{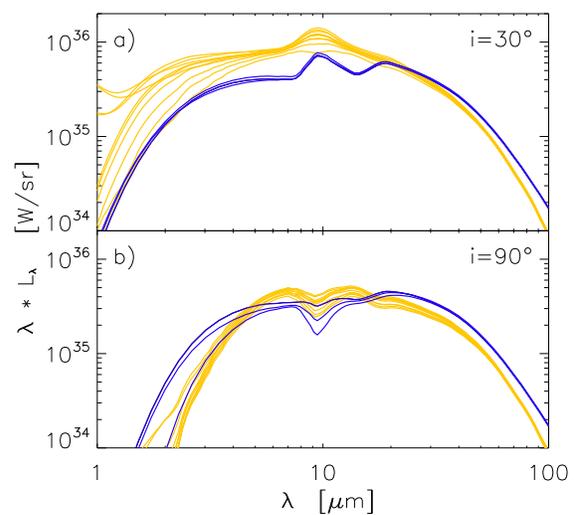}} 
  \caption[Comparison of our clumpy standard model with simulations by
           S.~F.~H\"onig]{Comparison 
           of our clumpy standard model and two other random realisations of the clump 
           distribution (blue lines) with
           simulations done by S.~F.~H\"onig
           (private communication, described in \citealp{Hoenig_06}), shown
           by the yellow lines, for 10 different random realisations of their
           model. The latter are scaled with a factor of 2.2 in order to give
           rough agreement between the two models (see text for further explanation).}   
  \label{fig:hoenig_comp} 
\end{figure}

A comparison of our clumpy standard model and two other random cloud
distributions with simulations by 
\citet{Hoenig_06} is shown in Fig.~\ref{fig:hoenig_comp}. 
They follow a different, multi-step approach: 2D radiative transfer
calculations of individual clouds at different positions and with various illumination 
patterns within the torus are carried out. 
In a second step, the SED of the total system is calculated.
The cloud distribution and parameters such as optical depth or size arise from an accretion 
scenario of self-gravitating clouds close to the shear limit \citep{Vollmer_04,Beckert_04}. 
The advantage of this approach is that
resolution problems can be overcome easily, as only 2D real radiative transfer 
calculations of single clumps are needed.
Characteristics for their modelling are small cloud sizes with very high optical depths 
in the inner part of the torus and a large number of clumps. 
For comparison, a cloud at the sublimation radius of their model has a radial size of 
$R_{\mathrm{cloud}}=0.02\,$pc with an optical depth of 
$\tau_{0.55\,\muup\mathrm{m}}^{\mathrm{clump}}\approx 250$. 
In our standard model, clouds at the sublimation radius are four times larger and possess
an optical depth of only $\tau_{0.55\,\muup\mathrm{m}}^{\mathrm{clump}}\approx 3$.
The large optical depth in the innermost part in combination with the large
number density there reduces the silicate feature significantly by shadowing 
with respect to their single clump
calculations. Their finding that the silicate emission 
feature can be reduced further by increasing 
the number density of clumps in the innermost part perfectly fits 
our explanation presented in Sect.\,\ref{sec:model_reduction}. 
Deviations between the two approaches (see Fig.~\ref{fig:hoenig_comp}) are mainly due to
the approximately eight times larger primary luminosity and the larger optical
depth, at least in the midplane of the \citet{Hoenig_06} modelling compared to
our standard model. Furthermore, 
in our simulations only dust reemission SEDs are shown. 
This leads to relatively higher fluxes at short
wavelengths compared to long wavelengths for the $i=30\degr$ case
(Fig.~\ref{fig:hoenig_comp}a) and to more extinction within the midplane and,
therefore, a shift of the Wien branch towards longer wavelengths in the edge-on
case (lower panel).

\section{MIDI interferometry}
\label{sec:MIDI_interferometry}

Even with the largest single-dish mid-infrared telescopes, it is impossible to directly resolve the dust torus
of the nearest Seyfert galaxies.  
Therefore, interferometric measurements are needed. Recently,
\citet{Jaffe_04} succeeded for the first time to resolve the dusty structure around an AGN in the 
mid-infrared wavelength range. In
this case, they probed the active nucleus of the nearby Seyfert\,2 galaxy NGC\,1068 with the help of the
MID-infrared interferometric Instrument (MIDI, \citealp{Leinert_03}). It is located at the European Southern 
Observatory's (ESO's) Very Large Telescope Interferometer (VLTI) laboratory on Cerro Paranal in Chile.
Its main objective is the coherent combination of the beams of 
two 8.2\,m diameter Unit Telescopes (UTs) in order to obtain structural properties of the observed objects 
at high angular resolution. A spatial resolution of up to $\lambda / (2\,B) \approx 10\,$mas 
at a wavelength of $\lambda=10\,\muup$m can be obtained 
for the largest possible separation of two Unit Telescopes of $B \approx 120\,$m.
Operating in the N-band ($8-13.5\,\muup$m), 
it is perfectly suited to detect thermal emission of dust in the innermost parts 
of nearby Seyfert galaxies. 
MIDI is designed as a classical Michelson interferometer. Being a two-element beam combining instrument, it
measures so-called visibility amplitudes. 
Visibility is defined as the ratio between the correlated flux and the total flux.
Its interpretation is not straightforward, since no direct image can be reconstructed.
Therefore, a model has to be assumed,
which can then be compared to the visibility data. 
\begin{figure}
  \centering
  \resizebox{\hsize}{!}{\includegraphics[]{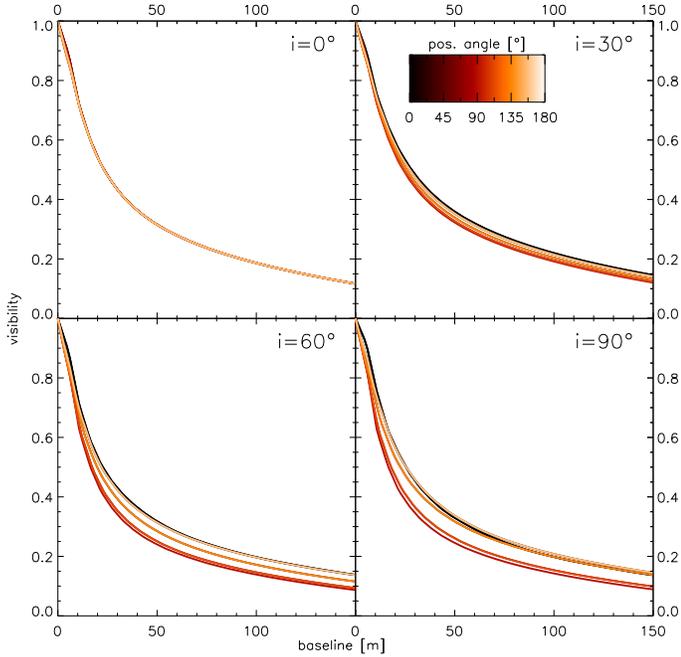}} 
  \caption[Visibilities of the continuous model]{Visibilities of our continuous standard model 
    at a wavelength of $12\,\muup$m
    plotted against the projected baseline length. 
    Colour of the visibility distributions 
    refers to different position angles of the projected baseline w.~r.~t.~the torus axis.
    Each panel shows a different inclination angle, as
    indicated in the upper right corner.}   
  \label{fig:vis_angles_cont} 
\end{figure}
MIDI works in dispersed mode, which means that visibilities for the whole wavelength range 
are derived. The dust emission is probed depending on the orientation of the projected baseline. 
Point-like objects result in a visibility of one, as the correlated flux equals the total flux. 
The more extended the object, the lower the visibility.
With the help of a density distribution, surface brightness distributions in the mid-infrared 
can be calculated by applying
a radiative transfer code. A Fourier transform of the brightness distribution 
then yields the visibility information, 
depending on the baseline orientation and length within the so-called {\it
U-V-plane} (or {\it Fourier-plane}).

The main goal of the following analysis is to investigate whether MIDI can distinguish
between clumpy and continuous torus models of the kind presented above.
Furthermore, we try to derive characteristic features of the respective models and
show a comparison to data obtained for the Circinus galaxy.

\subsection{Model visibilities}
\label{sec:model_visi}

\begin{figure}
  \centering
  \resizebox{\hsize}{!}{\includegraphics[]{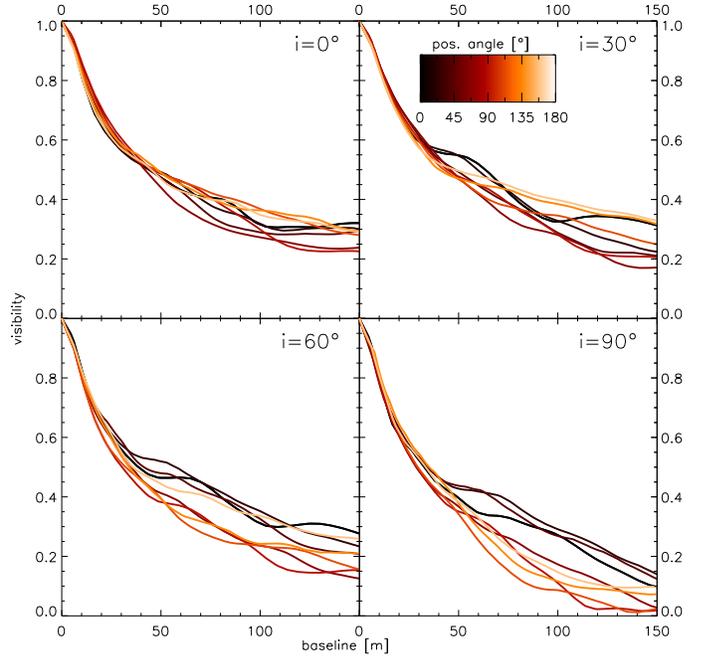}} 
  \caption[Visibilities of our clumpy standard model]{Visibilities of our clumpy standard model at a wavelength of $12\,\muup$m
    plotted against the projected baseline length. Colour of the visibility distributions 
    refers to different position angles of the baseline w.~r.~t.~the torus axis. 
    Each panel shows a different inclination angle, as
    indicated in the upper right corner.}   
  \label{fig:vis_angles_clumpy} 
\end{figure}

In Fig.~\ref{fig:vis_angles_cont},  
calculated visibilities for four inclinations of our continuous standard model at 
a wavelength of $\lambda$\,$=$\,$12\,\muup$m are
shown. Various orientations of the projected baseline are colour coded (the given 
position angle is counted anti-clockwise from the projected torus axis). 
Due to the axisymmetric setup, all lines
coincide for the face-on case. For all other inclination angles, visibilities
decrease until a position angle of $90\degr$ is reached and increase
symmetrically again. This means that the torus appears elongated
perpendicular to the torus axis at this wavelength.     
Fig.~\ref{fig:vis_angles_clumpy} shows the same study, but for the
corresponding clumpy model. The basic behaviour is the same, but the visibilities 
show fine structure
and the scatter is much greater, especially visible in the
comparison of the $i=0\degr$ cases.
Furthermore, while all of the curves of the continuous model monotonically 
decrease with baseline length, we see rising and falling values with increasing baseline
length for the same position angle in the clumpy case. In addition, for the continuous models,
curves do not intersect, in contrast to our clumpy models. 
However, to detect such fine structure in observed MIDI data, a very high
accuracy in the visibility measurements of the order of $\sigma_v \approx 0.02$ 
and a very dense sampling is required.

\begin{figure*}
  \centering
  \resizebox{0.75\hsize}{!}{\includegraphics[]{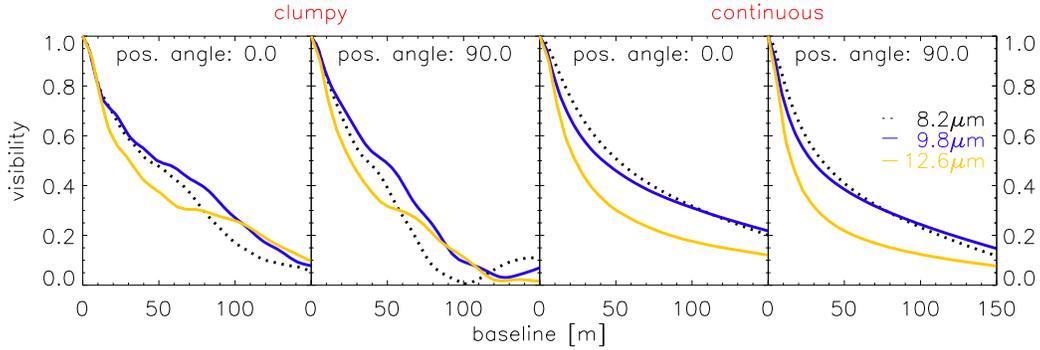}} 
  \caption[Comparison of visibilities of our standard clumpy and continuous
    model at different wavelengths]{Visibilities of our clumpy (first two panels) and continuous (last two
    panels) standard
    model at different wavelengths (colour coded), 
    plotted against the projected baseline length for the two position angles $0\degr$ (in
    torus axis direction) and $90\degr$ (along the midplane) for an
    edge-on view onto the torus ($i=90\degr$).}   
  \label{fig:vis_lam_clump} 
\end{figure*}

In Fig.~\ref{fig:vis_lam_clump}, the wavelength dependence of the visibilities
is shown. 
The first two panels represent the
case of the clumpy standard model and the third and fourth the continuous model.
Each of the two panels of the respective model visualises a different position angle (counted
anti-clockwise from the projected torus axis). 
An inclination angle of $90\degr$ is used in all panels.
Three different wavelengths are colour coded: $8.2\,\muup$m at the beginning of 
the MIDI-range (black dotted line), $9.8\,\muup$m within the silicate feature (blue) and $12.6\,\muup$m at
the end of the MIDI wavelength range (yellow), outside the silicate feature. 
While the continuous model results in smooth curves (see also Fig.~\ref{fig:vis_angles_cont}),
much fine structure is visible for the case of the clumpy model.  
The differences between the displayed wavelengths relative to the longest
wavelength are smaller for the clumpy models than in the continuous case.

\begin{figure}[t!]
  \centering
  \resizebox{\hsize}{!}{\includegraphics[]{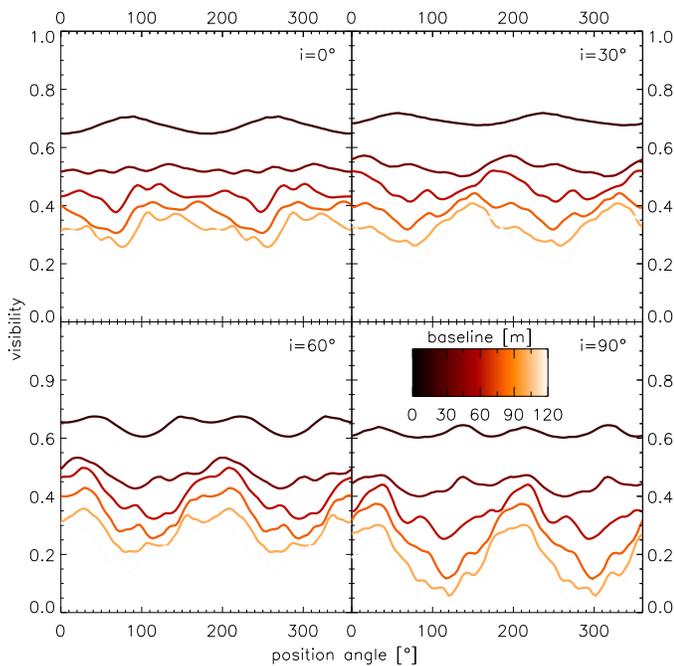}} 
  \caption[Visibilities of our clumpy standard model plotted against position angle]{Visibilities of our clumpy standard
    model at different inclination angles (as annotated in the upper right corner) 
    plotted against the position angle for various projected baseline lengths (colour coded)
    and a wavelength of $12\,\muup$m.}   
  \label{fig:vis_inc_posang} 
\end{figure}
 
Fig.~\ref{fig:vis_inc_posang} shows visibilities for our clumpy standard
model at $12\,\muup$m, plotted against the position angle (counter-clockwise from the
projected torus axis). Baselines are colour coded between $20\,$m and $100\,$m in steps of
$20\,$m. A longer baseline means that structures are better resolved, leading
to decreasing visibilities. For the case of inclination angles close to
edge-on, the visibility distribution changes from more or less flat to a
characteristic oscillating distribution at longer baselines (from $60\,$m
onwards) with minima around $100\degr$ and $300\degr$. This means that our
torus model seems to be more elongated within the equatorial plane and has the
smallest width along the projected torus axis at this wavelength. But this only applies for the innermost
part; the torus as a whole looks approximately spherically symmetric.         
At small inclination angles no such favoured size distribution  
is visible.

\subsection{Comparison with MIDI-data for the Circinus galaxy}
\label{sec:MIDI_comp}

\begin{table}[b!]
\caption[Parameters for the Circinus model]{Circinus model parameters: For an explanation of the parameters
  see Sect.\,\ref{sec:Model} and Table\,\ref{tab:model_param}. $M_{\mathrm{BH}}$ is the mass of the central 
  black hole \citep[from][]{Greenhill_03} and $L_{\mathrm{disk}}/L_{\mathrm{edd}}$ 
  is the Eddington luminosity ratio resulting for the assumed luminosity of the central source.
 }
\centering
\begin{tabular}{lcclc}
\hline
\hline
Parameter & Value & \hspace{0.5cm} & Parameter & Value \\
\hline
{\bf $R_{\mathrm{in}}$} & 0.6 pc & & {\bf $N_{\mathrm{clump}}$ } & 500 \\
$R_{\mathrm{out}}$ & 30 pc & & {\bf $\beta$}   &  1.0 \\
{\bf $\theta_{\mathrm{open}}$} & $65\degr$  & & {\bf $a_0$} &  0.2 pc\\
{\bf $\left<\tau_{9.7\,\muup\mathrm{m}}^{\mathrm{equ}}\right>_{\phi}$} & 3.9 
        &  & {\bf $\tau_{9.7\,\muup\mathrm{m}}^{\mathrm{clump}}$ } & 0.96 \\
{\bf $M_{\mathrm{BH}}$} & $1.7 \cdot 10^6\,M_{\sun}$  & & {\bf $L_{\mathrm{disk}}/L_{\mathrm{edd}}$} & 30\%\\ 
{\bf $\alpha$}  & -0.5 & &  & \\
\hline
\end{tabular}
\label{tab:Cir_param}
\end{table}

\begin{figure*}
 \begin{minipage}[b]{0.47\linewidth} 
  \resizebox{1.0\hsize}{!}{\includegraphics[]{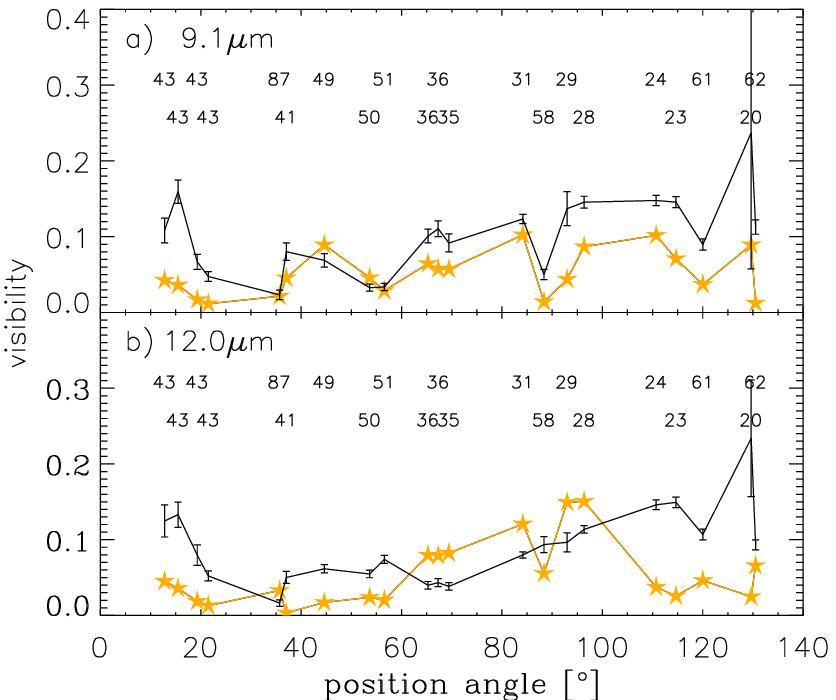}} 
  \caption[Comparison of model visibilities with data of the Circinus
    galaxy]{Comparison of 
    model visibilities (yellow stars and lines) for an azimuthal
    viewing angle of  
    $\phi=225\degr$ with MIDI observations for two different
    wavelengths. The baseline length for all data points is given above the data. 
    Data courtesy of \citet{Tristram_07}. \vspace{0.4cm}}   
  \label{fig:vis_circinus} 
 \end{minipage}
 \hfill 
 \begin{minipage}[b]{0.47\linewidth}
  \resizebox{1.0\hsize}{!}{\includegraphics[]{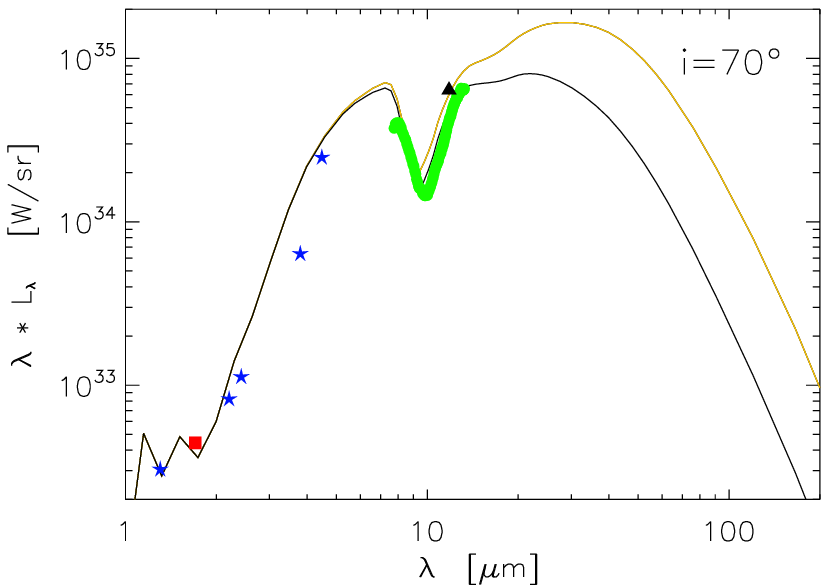}} 
  \caption[Comparison of SEDs of our clumpy Circinus model with high
   resolution data of 
   the Circinus galaxy]{Comparison of model SEDs with data for the Circinus galaxy. 
   Different symbols refer to various aperture radii:
   blue stars -- $0.38\arcsec$ (NACO), red rectangle -- $0.1\arcsec$ (HST/NICMOS) 
   and black triangle -- $1.0\arcsec$ (ESO/TIMMI2).
   Data compilation by \citealp{Prieto_04}. The thick green line shows 
   the total MIDI spectrum \citep{Tristram_07}.
   Our model (see model parameters in Table\,\ref{tab:Cir_param}) is calculated for an aperture radius
   of $0.4\arcsec$ (black line) and the total model space (yellow line). 
}   
  \label{fig:sed_circinus} 
 \end{minipage}
 \end{figure*}

Unfortunately, a fitting procedure involving a large parameter study is not possible 
with our current model, due to the very long 
computational times of the order of 30 to 40 hours per inclination angle (including calculation 
of the temperature distribution, the SED and surface brightness distribution). 
Therefore, we applied the following procedure: 
From our experience with modelling the SED of the Circinus galaxy with our previously used continuous 
{\it Turbulent Torus Models} (see \citealp{Schartmann_05}), 
we adopt the size of the object used there. Furthermore, we 
tried to stay as close to our clumpy standard model as possible (for the parameters of the 
clumpy standard model, compare to Table\,\ref{tab:model_param}) and copied the parameters $\alpha$, $\beta$ 
and $a_0$. The rest of the parameters were changed, in order to obtain the best possible adaptation 
to the data, within the investigated parameter range.  
The comparison of our current clumpy Circinus model as described above 
and in Table\,\ref{tab:Cir_param} (yellow stars) to
interferometric observations with MIDI \citep{Tristram_07} of the 
Circinus galaxy (black) is shown in  
Fig.~\ref{fig:vis_circinus}. 
In contrast to the presentation of continuous visibility curves above, 
single measurements of combinations of various baseline lengths and position angles are
displayed in this plot. Position angle now refers to the angle on the sky measured from north
in a counter-clockwise direction. The rotation axis of our simulated torus has a position angle of
approximately $-45\degr$
according to this definition.
The black numbers denote the length of the projected baseline (given in m) of the corresponding data point.  
From the approximate correspondence of the model values with the data, one can see that the 
size of the emitting region at the two wavelengths is reproduced
quite well. Most of the local extrema of the curve can be reproduced for the case of $9.1\,\muup$m. 
Larger deviations are visible for $\lambda = 12.0\,\muup$m. The good adaptation partly is due to
the changes in baseline length. Longer baselines naturally result in smaller visibilities, as we are probing
smaller and smaller structures (see also Fig.\,\ref{fig:vis_angles_clumpy}). 
Greater visibilities for shorter or equal baselines and similar position angle, 
therefore, have to be due to those curves in 
Fig.\,\ref{fig:vis_angles_clumpy} with increasing visibility with baseline or a very inhomogeneous distribution 
of dust with position angle. Both can be interpreted as signs of clumpiness.
The SED of the same Circinus model is plotted 
over current high resolution data in Fig.\,\ref{fig:sed_circinus}. 
The NIR (near-infrared) data points were obtained with the NACO camera at the VLT and corrected for foreground 
extinction by $A_{\mathrm{V}}=6\,$mag \citep{Prieto_04}. 
Different symbols refer to various aperture sizes (see figure caption). The thick green line 
shows the MIDI spectrum \citep{Tristram_07} and the black line is our Circinus model as 
discussed above for an aperture of $0.4\arcsec$ in radius; the yellow line denotes
the same model, but calculated for the whole simulated model space. 
Both modelled SEDs include the direct radiation of the central source (calculated with real Monte Carlo radiative 
transfer), which in these examples dominates over dust reemission for the small wavelength part from about 2 to $3\,\muup$m 
downwards and shows some noise, due to the low photon packet numbers used. 
In contradiction to our continuous Circinus model in
\citet{Schartmann_05}, enough nuclear radiation can be observed
in order to explain the turnover of the SED at small wavelengths and we do not
need to assume scattering by material (dust and electrons) within the torus
funnel. As can be seen from these figures, our model is able to qualitatively explain 
the SED as well as the visibility information. 
However, as we are not able to investigate the whole parameter range
of our models, we cannot exclude that a different parameter set can 
describe the data equally well. This degeneracy problem was already pointed out
by \citet{Galliano_03} for the case of SED fitting. Adding new clumpiness parameters will
even strengthen this degeneracy. On the other hand, adding more data such as more visibility
information will place more constraints and will weaken this problem.

\section{Conclusions}
\label{sec:conclusions}

In this paper, we implemented a new clumpy torus model in three
dimensions. For computational reasons, a wedge-like shaped disk
is used. In the discussion of our results, we place special emphasis on the
comparison with continuous models and their  
differentiation using interferometric observations, such as with MIDI. 

In \citet{Schartmann_05}, we had found that the SEDs of AGN
tori in the mid-infrared
wavelength range are mainly determined by the innermost part of the torus. 
With the presented clumpy torus models, this claim can be
further strengthened.
According to the new simulations, the silicate feature strength is mainly
determined by the number density and distribution, as well as the  
optical depth and size of the clumps in the inner region. With a 
sufficiently high optical depth
of the clouds in the inner part, shadowing effects become important, 
which hide the illuminated cloud surfaces from direct view
and, thereby, reduce the silicate feature in emission. At the same time,
enough lines of sight with low optical depth remain so that only weak
absorption features result for the edge-on case. 
Continuous models with special and unrealistic morphologies (like the
wedge-shaped tori used here) are also 
able to weaken the silicate emission feature for the face-on view when applying an 
anisotropic radiation characteristic, but fail to simultaneously 
account for moderate absorption features, when looking edge-on to the torus.

Due to the large clumps in our model, appreciable scatter in SEDs
for different random realisations of the torus are expected. 
A contrary effect is caused by the small optical depth of the single
clumps and also of many dust-free lines of sight towards the centre. 
Direct comparison between calculated interferometric visibilities for clumpy and the corresponding 
continuous models show that clumpy models naturally possess more
fine structure, which can partly be resolved by MIDI. 

We also showed that these kinds of models are able to qualitatively 
describe the available interferometric visibility and high resolution spectroscopic data of the 
Circinus galaxy at the same time. Currently, it is one of the best studied
Seyfert galaxies in terms of 
mid-infrared visibility measurements \citep{Tristram_07}.
The decreasing slope of the SED at short wavelengths 
can be described with our clumpy model, whereas it was at odds with the
continuous model described in \citet{Schartmann_05}.

\vspace{0.5cm}

\begin{acknowledgements}

We would like to thank the anonymous referee for comments, as well as
C.\,P.\,Dullemond for useful discussions and
S.\,F.\,H\"onig for providing some of his torus models for the 
comparison with our work. S.\,W.\,was supported by the German 
Research Foundation (DFG) through the Emmy Noether grant WO\,857/2.

\end{acknowledgements}

\bibliographystyle{aa}
\bibliography{astrings,literature}

\end{document}